# Terahertz PHASR Scanner with 2 kHz, 100 picosecond Time-Domain Trace Acquisition Rate and an Extended Field-of-View Based on a Heliostat Design

Zachery B. Harris and M. Hassan Arbab, *Member, IEEE*

*Abstract*— Recently, we introduced a Portable HAndheld Spectral Reflection (PHASR) Scanner to allow THz Time-Domain Spectroscopic (THz-TDS) imaging in clinical and industrial settings using a fiber-coupled and alignment-free telecentric beam steering design. The key limitations of the version 1.0 of the PHASR Scanner were its field-of-view and speed of time-domain trace acquisition. In this paper, we address these limitations by introducing a heliostat geometry for beam scanning to achieve an extended field-of-view, and by reconfiguring the ASynchronous OPtical Sampling (ASOPS) system to perform Electronically Controlled OPtical Sampling (ECOPS) measurements. The former change improved the deflection range of the beam, while also drastically reducing the coupling of the two scanning axes, the combination of which resulted in a larger than four-fold increase in the FOV area. The latter change significantly improves the acquisition speed and frequency domain performance simultaneously by improving measurement efficiency. To accomplish this, we characterized the non-linear time-axis sampling behavior of the electro-mechanical system in the ECOPS mode. We proposed methods to model and correct the non-linear time-axis distortions and tested the performance of the high-speed ECOPS trace acquisition. Therefore, here we introduce the PHASR Scanner version 2.0, which is capable of imaging a 40×27 mm$^2$ FOV with 2000 traces per second over a 100 picosecond TDS range. This new scanner represents a significant leap towards translating the THz-TDS technology from the lab bench to the bedside for real-time clinical imaging applications.

*Index Terms*—Terahertz Imaging, Terahertz Time-Domain Spectroscopy,

## I. INTRODUCTION

TERAHERTZ (THz) imaging has enjoyed many diverse potential applications such as art preservation [1], [2], security screening [3], [4], non-destructive testing [5]–[7], and biomedical analysis [8]–[12]. However, many of the available imaging devices are not field-deployable or rely on raster scanning of the sample or instrument for image formation. Current THz camera systems do not provide spectroscopic information from the sample [13]–[18]. Therefore, images obtained from today's THz cameras are not well suited for common techniques such as spectral "fingerprinting" [5], material parameter extraction (e.g. measuring refractive index) [8], [19], and analysis of scattering behavior [20]–[23]. In contrast, the time-domain spectroscopy (TDS) method can be used in these studies. Additionally, THz-TDS can provide structural information and sub-surface imaging based on time-of-flight analysis [24], [25]. Compressive sensing techniques allow for THz-TDS image formation using a stationary system, but its field of view is limited to the collimated beam width [26]–[29].

Portable THz spectroscopy has been demonstrated for single-point measurement using the battery-powered Micro-Z [30] and Mini-Z [31], [32] devices. Also, one-dimensional line scanning has been demonstrated using beam-steering along a single axis [33], [34]. However, in order to form an image, these devices would still need to be mechanically translated across the surface of a target. To address the need for portable full spectroscopic THz imaging devices, we developed the THz PHASR (Portable HAndheld Spectral Reflection) Scanner [35], [36]. This instrument acquired THz-TDS images over a 12×19 mm$^2$ field of view (FOV) using an f-θ lens and a mirror mounted in telecentric alignment on a motorized gimbal. An ASynchronous Optical Sampling (ASOPS) system was used to provide fast acquisition rates of 100 waveforms/s. Recently, this device has been demonstrated in the assessment and longitudinal monitoring of burn injuries in an in vivo porcine model [37]. However, the use of the scanner in our preclinical studies have highlighted two key limitations in the first version of the scanner: (i) the FOV, limited by the distortions inherent to its scanning geometry and the mechanical limits of the gimbal, and (ii) scanning speed, limited by the acquisition rate of the ASOPS technique. Solutions to these two limitations are crucial in translation of our technology from bench to the bedside in the upcoming pilot human studies.

Here, we present the new THz PHASR Scanner 2.0, shown in Fig. 1(a). To increase the FOV, we redesigned the beam steering geometry based on a heliostat configuration, which

Research reported in this publication was supported by the National Institute of General Medical Sciences of the National Institutes of Health under award number R01GM112693. *(Corresponding author: Hassan Arbab).*
The authors are with the Department of Biomedical Engineering, Stony Brook University, Stony Brook, New York 11794, USA (e-mail: zachery.harris@stonybrook.edu; hassan.arbab@stonybrook.edu).

See supplementary material for the video of a 27×27 mm$^2$ FOV scan.

Color versions of one or more of the figures in this article are available online at http://ieeexplore.ieee.org

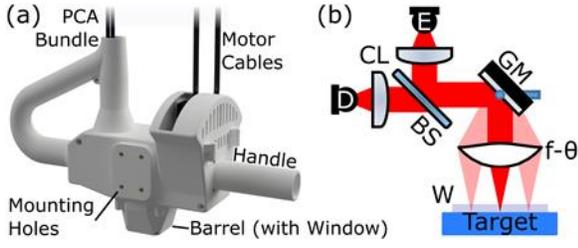

**Fig. 1.** (a) The PHASR 2.0 scanner. (b) Our telecentric imaging system with photoconductive antenna (PCA) emitter and detector, E and D, respectively, each paired with a collimating/focusing lens, CL. A silicon beam splitter, BS, directs the beam to the collocated section containing the gimbaled beam-steering mirror, GM, and f-θ imaging lens. An optional imaging window, W, is shown at the target plane.

additionally eliminated the distortions due to the intercoupling of the scanning axes in the gimballed motors of the PHASR Scanner 1.0. To increase the speed of the TDS trace acquisitions, we have adapted our existing ASOPS electronic hardware to perform Electronically Controlled OPtical Sampling (ECOPS) instead. These changes produce a new scanner with a large, 40×27 mm² FOV and capable of recording 2000 waveforms per second, representing a 20-fold increase in acquisition speed.

## II. Design of the PHASR 2.0 Scanner

The schematic of a general telecentric THz-TDS imager is shown in Fig. 1(b). The terahertz light is generated by a commercial fiber-coupled photoconductive antenna (PCA), collimated, and then directed through a beam splitter. The beam is steered across a custom high-density polyethylene (HDPE) f-θ lens by a gimballed mirror located at the lens' rear focal point, thus creating a telecentric configuration. In this design, the lens maintains a normal incidence angle on the target, a flat focal surface plane, a constant focal spot size, and constant optical path length for all positions within the FOV [38]. The normal incidence and flat focal plane mean that the reflected beam is collocated with the incident beam, returning by the same path to the beam splitter where it is directed towards the detector PCA. Optionally, an imaging window can be used at the focal plane to flatten soft targets and allow for self-calibration reference measurements using the air-window interface reflections [19].

The simplified model of a generalized gimbal in Fig. 2(a) shows how rotations about the outer gimbal axis (blue, $\alpha$) change the orientation of the other axis (red, $\beta$). In our previous design, we used a commercial gimbal unit with ±7° deflection in each axis. Due to the specific gimbal architecture, the unit was mounted at a 45° angle as shown in Fig. 2(b). In this arrangement, the angle between the elevation axis and the incident beam, $b_{\text{in}}$, is dependent on the orientation of the azimuthal axis. Note that, at the default position of the azimuthal motor, $\alpha = 0°$ shown in Fig. 2, the elevation axis, $\beta$, is perpendicular to the incident beam, whereas if the azimuthal axis rotates to its maximum range, 7°, the angle between the elevation axis and the beam would be about 5°. As a result of this varying angle, movement directly along the horizontal, $x$, or vertical, $y$, scanning directions requires contribution from both motors. In other words, the two gimbal axes are intercoupled. Fig. 2(c) shows the coordinate mapping from the motor angles to the position of the focused beam for this design, derived in detail in [35], which only provided a 12×19 mm² FOV.

To improve this range, we have redesigned the mirror gimbal layout as shown in Fig. 3(a). Inspired by heliostats, instruments used in astronomy to reflect light from the sun as it moves through the sky to a fixed point, we have adapted the scanning mechanism's orientation to reduce the axial coupling [39]. Instead of a single off-the-shelf gimbal, a pair of motors were stacked in a "daisy-chained" configuration. A rotation stage controlling the azimuthal axis is fastened directly to the scanner housing. The elevation angle is controlled by a motorized goniometer attached to the rotation stage. A 3D printed mirror mount biased by 45° about the elevation axis is used to properly locate the mirror for scanning. The model gimbal shown in Fig. 3(b) demonstrates this orientation. Note again the effect that rotating about the azimuthal axis has on the angle between the incident beam and the elevation axis. In this design, the outer azimuthal axis is collinear with the incident beam and as such, the elevation axis remains perpendicular to the incident beam at any azimuthal position. This provides the larger and significantly more rectilinear FOV shown in Fig. 3(c). For comparison, the outline of the PHASR Scanner 1.0 FOV is shown by the black dashed line. The vertical scan range, limited by the ±10° travel of the goniometer, is approximately 27 mm at the center, expanding slightly at larger horizontal positions. The color within the scanning area in Fig. 3(c) shows the simulated normalized power at the target calculated via ray-tracing. The circular profile shows how the primary limiting factor of the horizontal scan range is the diameter of the f-θ lens which provides approximately a 40-mm range.

### A. Heliostat Beam Scanning Algorithm

To demonstrate the decoupling of the imaging axes of rotation in the heliostat design, we derive the scanning coordinate system from the axial deflections. We define the $z$-axis to be aligned antiparallel with the optic axis of the f-θ lens, and the $x$- and $y$-axes as shown in Fig. 3(a). A vector perpendicular to the face of the mirror then has the direction

$$\hat{m} = \begin{pmatrix} \cos(\beta - 45°)\sin(\alpha) \\ \sin(\beta - 45°) \\ -\cos(\beta - 45°)\cos(\alpha) \end{pmatrix}, \quad (1)$$

where $\alpha$ and $\beta$ are the angles rotated by the azimuthal and elevation motors, respectively, and defined in Fig. 3(b). That is, the mirror points in a direction corresponding to a simple spherical coordinate system with azimuthal angle about the $y$-axis and elevation measured in either direction from the $xz$-plane. The collimated THz beam is then described by the incident and reflected vectors,

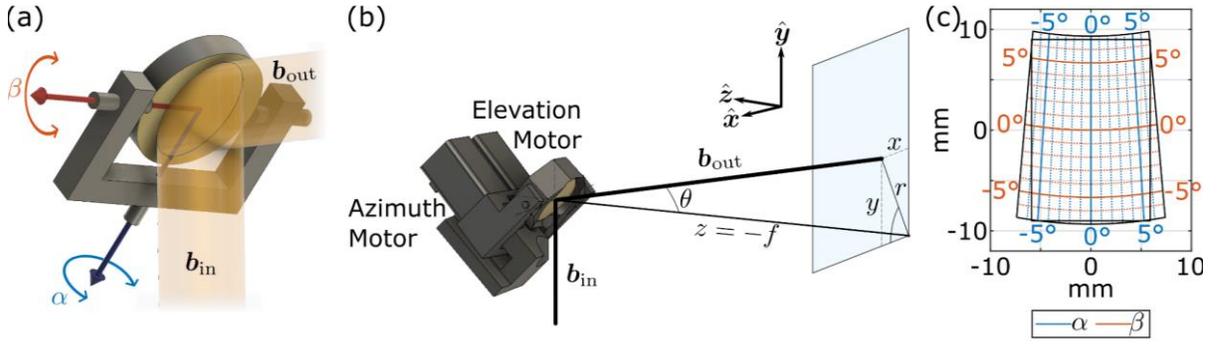

**Fig. 2.** (a) Geometry of the beam steering in PHASR Scanner 1.0. (b) Simplified representation of the gimballed mirror in PHASR 1.0. (c) Resultant scanning pattern from this geometry. Blue and red grids represent coordinates of angular deflection of the scanning mirror, α and β, about its azimuthal and elevation axes, respectively.

$$\widehat{\boldsymbol{b}}_{\text{in}} = \begin{pmatrix} 0 \\ 1 \\ 0 \end{pmatrix}, \qquad \boldsymbol{b}_{\text{out}} = \begin{pmatrix} x \\ y \\ -f \end{pmatrix}, \quad (2)$$

respectively, where $(x, y)$ is the location of the beam at the lens plane. Reflecting $-\boldsymbol{b}_{\text{in}}$ about $\boldsymbol{m}$ to get the direction of $\boldsymbol{b}_{\text{out}}$ and scaling such that the $z$ coordinate is equal to $-f$, so that $x$- and $y$-coordinates represent the location at the lens plane, it can be seen that

$$\boldsymbol{b}_{\text{out}} = \begin{pmatrix} f \tan(\alpha) \\ f \sec(\alpha) \tan(2\beta) \\ -f \end{pmatrix} = f \sec(\alpha) \begin{pmatrix} f \sin(\alpha) \\ \tan(2\beta) \\ -\cos(\alpha) \end{pmatrix}. \quad (3)$$

Equation (3) shows that at the lens plane, the $x$-coordinate is only dependent on $f$ and $\alpha$, and the value $h = f \sec \alpha = \sqrt{x^2 + z^2}$ is the distance to the lens plane at azimuthal angle $\alpha$. Similarly, the $y$-coordinate is only dependent on $h$ and $\beta$. That is, the angle within the $xz$-plane is determined only by $\alpha$ and the angle away from the $xz$-plane is determined only by $\beta$.

The basic scanning algorithm for this design is then as follows. The face of the mirror must point in the direction bisecting the incident and reflected beams:

$$\boldsymbol{m} = \|\widehat{\boldsymbol{b}}_{\text{in}}\| \boldsymbol{b}_{\text{out}} + \|\boldsymbol{b}_{\text{out}}\|(-\widehat{\boldsymbol{b}}_{\text{in}}) = \begin{pmatrix} m_x \\ m_y \\ m_z \end{pmatrix}. \quad (4)$$

The axes of the gimbal must rotate to

$$\alpha = \arctan\left(\frac{m_x}{-m_z}\right), \qquad \beta = \arctan\left(\frac{m_y}{\sqrt{m_x^2 + m_z^2}}\right) + 45°. \quad (5)$$

Our previously demonstrated linear correction can be applied to account for slight deviations due to the f-θ lens [35]. This method is general for heliostat scanning over f-θ lenses. Variations on this design using different lenses might provide better performance for different applications, albeit with some tradeoffs. For instance, a lens with a larger focal length provides a greater field of view for the same angular range of travel at the gimbal, but at the cost of a larger spot size at the focus, reducing the spatial resolution of the device. The particular f-θ lens used here has been developed to suit the needs of a compact scanner. Thus, the vertical scan range of the PHASR Scanner 2.0, limited by the ±10° travel of the goniometer, is approximately 27 mm at the center, whereas the horizontal scan range is limited only by the lens area (e.g., here to approximately 40 mm).

### III. ECOPS MEASUREMENTS USING ASOPS HARDWARE

The imaging rate of our previous scanner was limited by the measurement speed of the commercial ASOPS system used for generation and detection of THz pulses. Although faster than using a mechanical delay line, the THz-TDS acquisition rate was slower than speeds provided by the ECOPS technique. ECOPS trace acquisition rates of 2.5 kHz [40], 8 kHz [41] and even as high as 60 kHz [42] have been demonstrated, though with THz time-axis ranges limited to less than about 20 ps at those speeds. Imaging systems using ECOPS technique have been reported with operating speeds of 1000 trace/s [43], [44]. Also, point measurements of sample layer thickness have been acquired at 1600 Hz rates with 200 ps of range [45]. Both ASOPS and ECOPS use two femtosecond lasers to respectively generate and sample the temporal waveform of the THz electric fields. These two techniques are described in detail elsewhere [46], [47], however, an overview is provided here and illustrated conceptually in Fig. 4.

In both techniques, the difference in repetition rate of the two lasers causes the sampling laser to progressively record sequential THz pulses in time, building a representative time-domain acquisition. Here, we call the laser generating the THz pulses "Laser A," which has a constant repetition frequency $f_{\text{rep}}$ and will thus produce THz pulses separated by a period of $T_{\text{rep}} = 1/f_{\text{rep}}$. The laser sampling the THz pulses is called "Laser B" and has its repetition frequency set to $f_{\text{rep}} - \Delta f$, where $\Delta f$ is small compared to $f_{\text{rep}}$. In general, the difference frequency $\Delta f$ can be set by the user and is dependent on time, t. As a result, the THz pulse samples occur at a period of $1/(f_{\text{rep}} - \Delta f)$. Assuming no variation in the beam path, each of these THz pulses is essentially identical at the detector so the different repetition periods of the two lasers mean that, in comparison to the previous sampling location of the THz pulse, each subsequent sample will be delayed by [46], [47],

$$\Delta \tau(t) = \frac{1}{f_{\text{rep}} - \Delta f(t)} - \frac{1}{f_{\text{rep}}} \approx \frac{\Delta f(t)}{f_{\text{rep}}^2}, \quad (6)$$

where values of $\tau$ refer to the effective time axis intervals of the THz pulse and are usually in picoseconds. The actual sampling interval in lab time, $t$, is $T_{\text{rep}} = 1/f_{\text{rep}}$ and is equal to the period of the femtosecond laser pulses. Each successive pulse from Laser B will sample the corresponding THz pulse generated by Laser A at a time-point shifted by $\Delta\tau$. Starting at an arbitrary sampling time $\tau(0) = \Delta\tau_0$ at time $t = 0$, the effective sample time at $t$ is given by:

$$\tau(t) = \tau_0 + \sum_{n=0}^{\lfloor t \times f_{\text{rep}} \rfloor} \frac{\Delta f(n/f_{\text{rep}})}{f_{\text{rep}}} \times \frac{1}{f_{\text{rep}}} \quad (7)$$

where $n = \lfloor t \times f_{\text{rep}} \rfloor$ is the integer number of laser pulses that have occurred since $t = 0$. The separated factor of $1/f_{\text{rep}}$ in (7) emphasizes the fact that this is, in essence, a Riemann sum with its step size defined by the repetition interval of the laser. Thus, the transform between time and the sampling location can be approximated by the integral of the difference frequency over time.

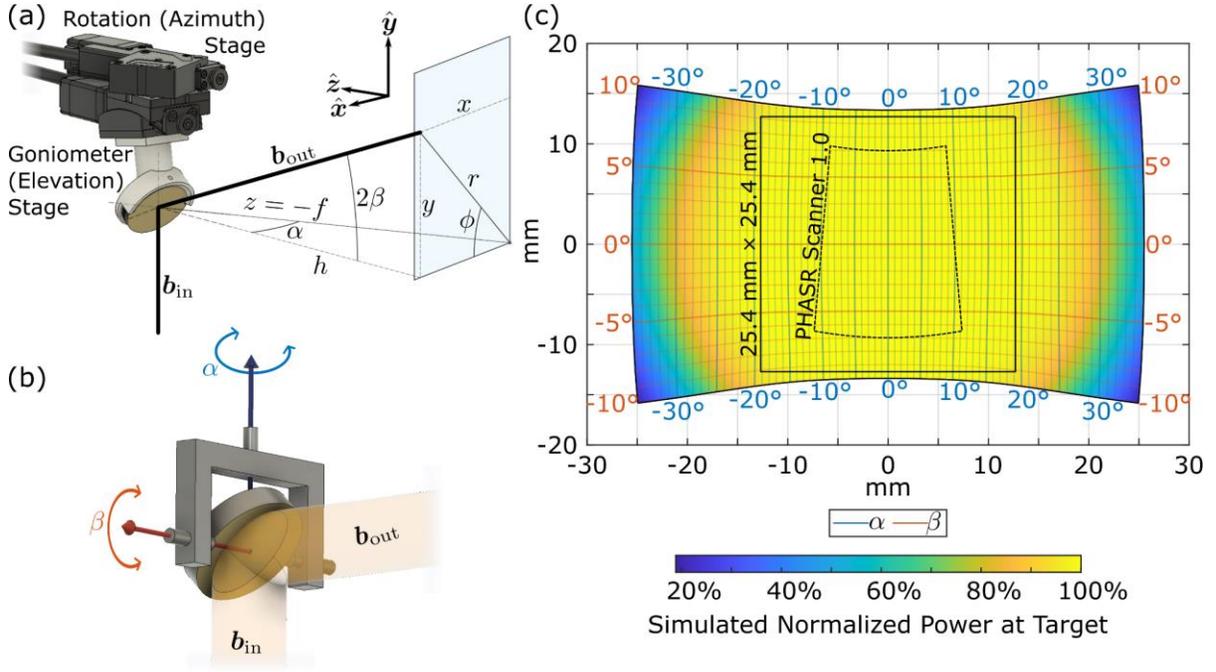

**Fig. 3.** (a) Geometry of the PHASR Scanner 2.0 beam steering. (b) Simplified representation of the gimballed mirror in PHASR 2.0 showing the azimuthal axis, in blue, aligned with the incident beam and elevation axis, in red, perpendicular to it. (c) Resultant scanning pattern from this geometry. Blue and red grids represent coordinates of the angular deflection of the scanning mirror, α and β about its azimuthal and elevation axes, respectively. The dashed black line shows the FOV accessible with the previous version of the scanner and the solid black line shows the typical scanning area of 25.4×25.4 mm2 (1×1 in.2). The color scale shows the normalized incident power at the target as determined by ray-tracing simulation.

In ASOPS measurements, illustrated in blue in Fig. 4, $\Delta f$ is kept constant and $\tau$ will increase by the same amount per pulse. The direction in which the sampling progresses depends on which laser has a higher repetition rate, i.e., it depends on the sign of $\Delta f$. If Laser B has a lower repetition frequency, as depicted in blue in Fig. 4(a), the sampling can be said to be in the "forward" direction as each subsequent sample is associated with a later time in the THz signal, as shown in Fig. 4(b). If the frequency of Laser B is higher, the sampling will occur in the opposite, "backwards," direction. For ASOPS measurement in either direction, after one full period of the difference frequency, $1/|\Delta f|$, the accumulated sample time will equal that of a full period of the laser repetition, that is, $\tau(1/|\Delta f|) - \tau_0 = T_{\text{rep}}$, and thus samples covering the full THz pulse will have been acquired. To improve the signal to noise ratio, SNR, multiple sequential acquisitions are then typically averaged to build a single THz-TDS trace, so the total time per trace is the number of averages multiplied by $1/|\Delta f|$.

However, ASOPS measurements are not time-efficient because in every acquisition event the entire $T_{\text{rep}}$ on the order of 10 ns, is recorded but only the relevant THz-TDS measurements range, typically on the order of 100s of ps, is retained. This effect is illustrated in Fig. 4(b) by the range between the dashed lines. Thus, the majority of the period of ASOPS is spent sampling timepoints outside of the range of interest. ECOPS improves the measurement speed by only sampling a small range of interest.

The ECOPS technique can be understood as ASOPS measurement with an alternating $\Delta f$, such as the illustrations shown in Fig. 4(a). As a result of the modulated $\Delta f$ value, instead of sampling the entire $1/|\Delta f|$ period, Laser B repetitively samples in both forward and backward directions over only a small section of the available period as shown in red in Fig. 4(b). If the frequency of the modulation is $f_M$, single-shot THz-TDS traces can be acquired at $2f_M$ since data can be recorded in both directions. However, unlike ASOPS, which can sample any-sized section of the entire $T_{\text{rep}}$ period of the THz waveform, ECOPS's measurement range, $T_{\text{THz}}$, is linked to the speed through both $\Delta f$ and $f_M$. Thus, while the THz acquisition window and its starting point, $\tau_0$, can be adjusted in ASOPS by simply recording a different section of the $1/|\Delta f|$ period, for ECOPS this requires more coordinated adjustment of the modulation parameters.

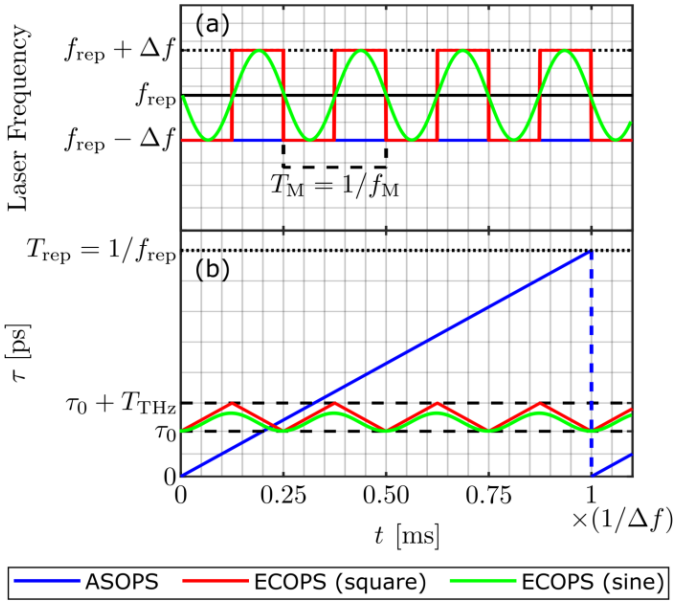

**Fig. 4.** Conceptual depiction of the difference between the ASOPS (blue lines) and the ECOPS techniques using square (red lines) and sinusoidal (green lines) modulation. (a) the repetition rate of the two lasers, where Laser A is represented by the solid black line for both ASOPS and ECOPS. The repetition rate of Laser B is constant in ASOPS (blue), whereas it is modulated in ECOPS with a square (red) or sinusoidal (green) function. (b) The time domain sampling instances, as given by Eq. (7), are depicted for both techniques. Dashed black lines indicate the portion of interest in the time domain sampling window in a typical THz-TDS measurement. The time axis, shared between both (a) and (b), shows one $1/|\Delta f|$ sampling period, i.e., the time required to record a single ASOPS trace, in laboratory time. In practice, the magnitude of $\Delta f$ is not the same between the two techniques, and a sinusoidal function of much higher frequency is used to drive ECOPS.

Despite these differences, the two techniques can be implemented using the same equipment. The variation in $\Delta f$ can be created by taking advantage of the existing ASOPS hardware. The repetition rate of each of the lasers is controlled by the length of the laser cavity using stepper motors and piezoelectric actuators for coarse and fine adjustment of the laser cavity. A feedback system monitors the pulse rates and compares them to a reference oscillator and then adjusts the cavity length accordingly in real time. In our system, the reference for Laser B is generated by a Keysight 33500B Series waveform generator, allowing a user to select different values of $\Delta f$ for different speeds of ASOPS. ECOPS operation is achieved by modulating the frequency of this reference and relying on the same feedback system to correctly adapt the repetition rate.

Practical hardware limitations mean that if the target $\Delta f$ is modulated too aggressively the system introduces errors in the repetition rate or will lose phase-locking between the two lasers entirely. For this reason, we found it preferable to use sinusoidal modulation, illustrated by the green traces in Fig. 4, which more smoothly varies the difference frequency than a square wave, shown in red. Likewise, to maintain phase-locking in our system we must use a peak ECOPS $\Delta f$ smaller than the typical $\Delta f$ used in ASOPS. However, the varying magnitude of the difference frequency produced by a sinusoidal modulation results in a time-dependent sampling rate of the THz time-domain demonstrated by the curved green trace in Fig. 4(b). This is further complicated by the imperfect system response to the modulated reference frequency. The combined effects of these present transiently as timing drift and persistently as a distortion to the expected sampling rate of the THz signal, both of which must be accounted for. As a shorthand, we will specify the ECOPS modulation parameters by the modulation frequency and peak frequency differences. For example, "$f_M = 1000$ Hz, $\Delta f = \pm 32$ Hz" should be understood as a nominal modulation of the form $\Delta f = (32 \text{ Hz})\sin(2\pi t(1000 \text{ Hz}))$. In contrast, ASOPS measurements, which have a constant difference frequency, will be indicated just by that value, e.g., $\Delta f = 100$ Hz.

*A. Timing drift*

Drift occurs if the system responds differently to the two forward and backward directions of time-domain scanning, e.g., through a hysteresis in the piezo. In that case, the two directions will cover different amounts of $\tau$, leading to an apparent drift of the THz signal due to the drift of the ECOPS sampling range. Left unchecked, this drift will quickly cause the region of interest to shift out of the scanning window. To counteract this effect, a small offset (typically on the order of tens of mHz) to the base repetition rate of laser B, $f_B$, is required to bias the modulation by the same amount opposite to the drift. However, in our system, using a constant offset value is insufficient as the drift varies over time, as much as 5 ps/s within minutes. To address this issue, we have implemented a state control model for real-time drift compensation described by the flowchart in Fig. 5. In order to properly track the drift, we must find and lock on to a feature (such as the peak of a THz pulse) known to be stationary. For rough-surface or malleable targets such as liquid or skin, the flat imaging window, shown in Fig. 1 provides an ideal reflection reference from the air-window interface. Since the f-θ lens provides a constant phase at its focus over the entire planar field of view [38], there is no additional compensation for scanning location needed. The difference of the current apparent time location of this feature, $\tau_{\text{Curr}}$, from the location measured some amount of time, $\Delta t$, previously, $\tau_{\text{Prev}}$, provides the drift, $d$, of the ECOPS time window. A small adjustment to the frequency offset is made to compensate any time the window drift is above a certain threshold, $d_{\text{Max}}$, in either direction. Once engaged, this compensation actively counters the majority of the drift in real-time. In this paper, we classify any remaining variation as jitter, which contributes to the measurement noise, and will be discussed in section IV(A).

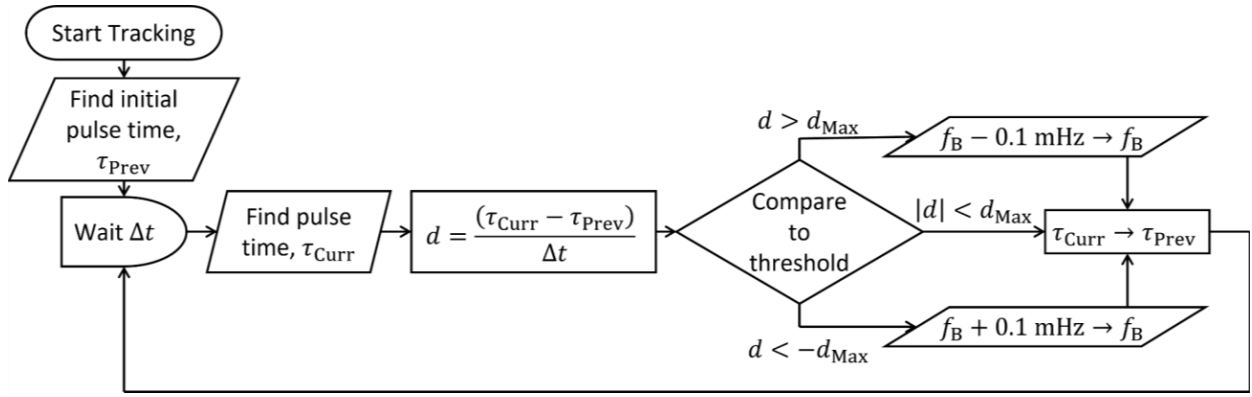

**Fig. 5.** Flowchart of state model drift compensation algorithm. Drift, $d$, is calculated as the change in measured feature location ($\tau_{\text{Curr}} - \tau_{\text{Prev}}$) over some amount of time, $\Delta t$. If the magnitude of $d$ is greater than a threshold, $d_{\text{Max}}$, then a small corresponding correction is made to the base repetition rate of Laser B, $f_B$.

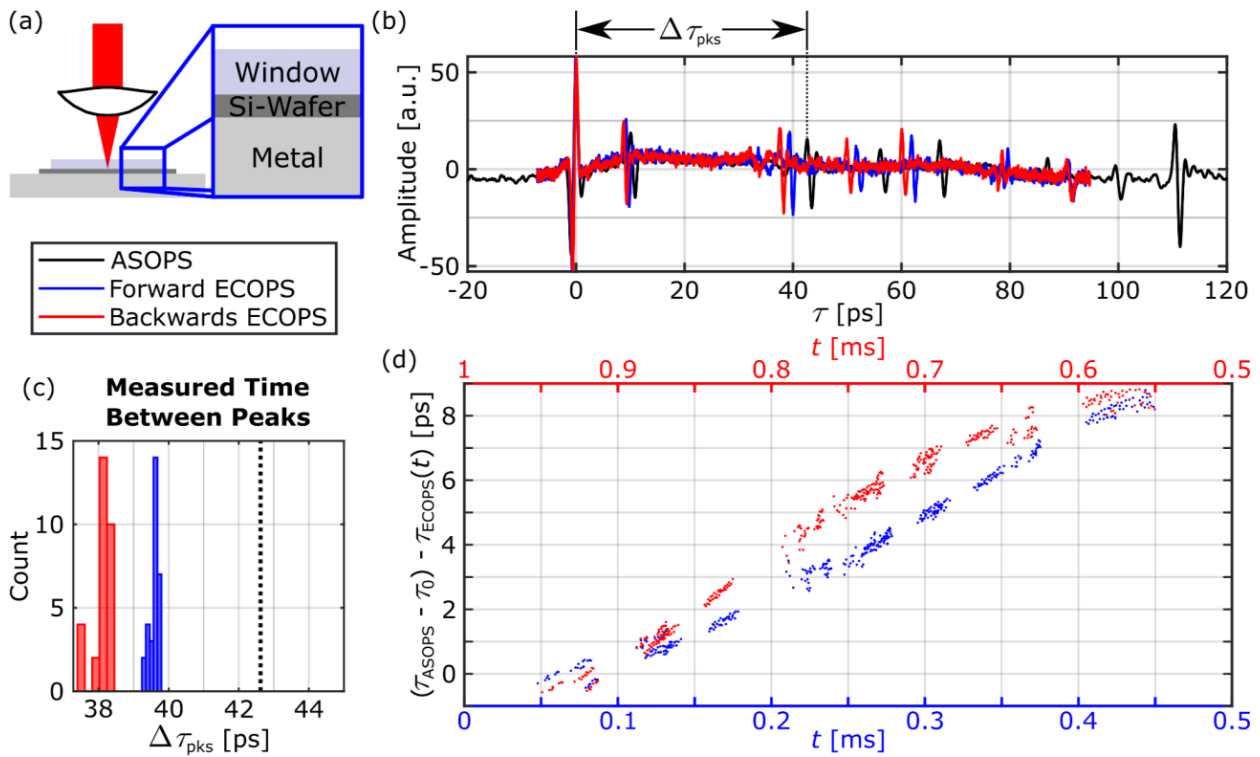

**Fig. 6.** (a) Illustration of the multi-layer reference target used for providing constant sampling points in the time-domain. (b) Representative time-domain signals when using the ECOPS time axis ($f_M = 1000$ Hz, $\Delta f = \pm 32$ Hz, 50 avg.) in the forward (blue) and backward (red) directions compared to the reference ASOPS signal (black) of the same target ($\Delta f = 100$ Hz, 100 avg.). (c) Distribution of the measured delay between the two pulses marked in (b) over 30 traces from each of the ECOPS directions with black dotted line showing the ASOPS value. (d) Comparison of the difference between the measured ASOPS and ECOPS locations (adjusted for different $\tau_0$ values) of the time-domain reflection peaks in the forward (blue, lower axis) and backward (red, upper axis) for all 30 ECOPS datasets.

*B. Nonlinear time-domain sampling and its correction*

In addition to the drift described in the previous section, other time-axis distortions are present as a result of the electronic and mechanical systems response to the frequency modulation. For example, any inaccuracy of the modulation electronics, mechanical responses of the piezos to the electronic waveform, or drift correction will modify the modulation from the nominal sinusoid and thus deviate the sampling from the expected points. The extent of this distortion is illustrated by measurements from a multi-layer reference target, shown in Fig. 6(a). TDS point measurements of a thin wafer of silicon, sandwiched between the imaging window and metallic back layer, provides many distinct pulses due to the Fabry-Perot reflections. The relative timing of these "landmark" features allow for simple comparison between ECOPS signals and an ASOPS reference measurement of the same location on the multi-layer sample. Figure 6(b) illustrates the ECOPS time axes generated by Eq. (7), $\tau_{ECOPS}(t)$, using the expected modulation function. This basic model does not result in the correct time axes and notably the timing error is not the same for both directions of the ECOPS sampling. For instance, note the interval labeled by $\Delta \tau_{pks}$ in Fig. 6(b). As measured by the ASOPS reference, the later peak arrives 42.62 ps after the first. As shown in Fig. 6(c), this difference is consistently underestimated by the ECOPS signals. Figure 6(d) shows the time-dependent difference in the ECOPS location of the landmark peaks from the same observed ASOPS locations, $\tau_{ASOPS}$, offset by the different $\tau_0$ values for each trace. If the ECOPS measurement perfectly reproduced the ASOPS signal, these points would fall at 0 for all $t$. Instead, the non-zero slopes of the two sets indicate that the scaling provided by Eq. (7) didn't capture the dynamic response of the electrical and mechanical hardware. The differences in the response to each direction of modulation is also made clear by plotting the forward and backwards ECOPS measurement using red and blue data points and corresponding time axes.

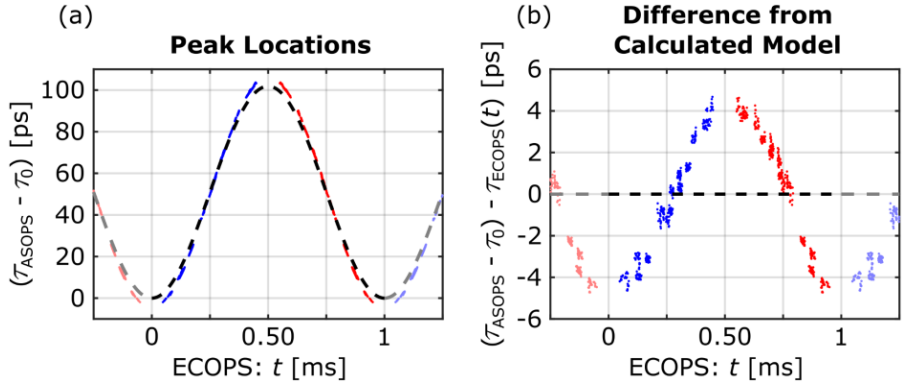

**Fig. 7.** (a) Correspondence of the location of the time domain peaks between the ECOPS forward (blue) and backward (red) directions and the ASOPS signals compared to the results calculated from Eq. (7) assuming that $(\tau_{ASOPS} - \tau_0) = \tau_{ECOPS}(t)$ (black dashed line). The data presented here is the same as those shown in Fig. 6 ($f_M = 1000$ Hz, $\Delta f = \pm 32$ Hz, 50 avg.) (b) The difference from the calculated model, more clearly showing the asymmetry of the trend between the ECOPS forward and backward directions. The peak locations in (a) and (b) were collected from 30 separate time-domain acquisitions.

The full extent of the timing error can be seen in Fig. 7(a) and 7(b), which compares the expected time axis function calculated from Eq. (7), black dashed line, to the actual corresponding ECOPS time-domain peaks in both forward (blue) and backwards (red) directions. The faded copies of the data points on either side of the 0 to $1/f_M$ period provide context to the behavior of the model near the start and end of each measurement cycle. The trend shown in Fig. 7(b) results in time-domain error values of up to 5 ps.

This method of comparison provides an approach for a simple time-axis correction using empirical measurements of the Fabry-Perot reflections in the time-domain signals. A similar method was used for finding the time-axis scaling factor in [41]. Since the reflections are distinct and deterministic events in the time domain, they can be used for discrete sampling of a transformation from $t$ to $\tau$. In other words, a function describing the transformation between the ECOPS locations of the "landmark" features, such as the peaks, and the corresponding features in an ASOPS acquisition provides time-axis calibration. We can approximate this transform for an individual acquisition using a polynomial equation of order $N$ given by,

$$\tau(t) - \tau_0 = \sum_{P=1}^{N} C_P t^P, \quad (8)$$

where $C_P$ are the coefficients of the $P^{th}$ polynomial term. If we label the landmark features $a, b, c, ...$ and associate their ECOPS locations in lab time: $t_a, t_b, t_c, ...$ with their ASOPS time-locations: $\tau_a, \tau_b, \tau_c, ...$ then the set of $C_P$ values and $\tau_0$ can be found using a least-squares fitting algorithm produced by numerically solving the matrix equation, given by

$$\begin{bmatrix} (t_a)^N & (t_a)^{N-1} & \cdots & t_a & 1 \\ (t_b)^N & (t_b)^{N-1} & \cdots & t_b & 1 \\ (t_c)^N & (t_c)^{N-1} & \cdots & t_c & 1 \\ \vdots & \vdots & \vdots & \vdots & \vdots \end{bmatrix} \begin{bmatrix} C_N \\ C_{N-1} \\ \vdots \\ C_1 \\ \tau_0 \end{bmatrix} = \begin{bmatrix} \tau_a \\ \tau_b \\ \tau_c \\ \vdots \end{bmatrix}. \quad (9)$$

The initial sampling point, $\tau_0$, depends on the starting point of the window thus in general will be different for each acquisition. The polynomial coefficient terms, $C_P$, model the shape of the ECOPS time sampling and, excluding jitter, are expected to be the same for each acquisition.

The accuracy of this approximation will be limited by the number and distribution of sampling points used to generate the fit. We further increase the number of reference points by using multiple acquisitions with different $\tau_0$ values, and thus different time window locations. In order to find the correct coefficients for all acquisitions, we fit a system of equations, in which the constant terms (0$^{th}$ order) are unique to acquisitions of the different time windows, but non-constant (1$^{st}$ and greater order) polynomial terms remain the same. That is, for $M$ different ECOPS acquisitions, we extract the time-locations, $t_{i,m}$ and $\tau_{i,m}$, where $i = a, b, c, ...$ and $m = 1, 2, ..., M$, for all "landmarks" time sampling locations. The system of equations can be represented by the matrix equation given by,

$$\begin{bmatrix} (t_{a,1})^N & (t_{a,1})^{N-1} & \cdots & t_{a,1} & 1 & 0 & 0 & \cdots & 0 \\ (t_{b,1})^N & (t_{b,1})^{N-1} & \cdots & t_{b,1} & 1 & 0 & 0 & \cdots & 0 \\ (t_{c,1})^N & (t_{c,1})^{N-1} & \cdots & t_{c,1} & 1 & 0 & 0 & \cdots & 0 \\ \vdots & \vdots & \ddots & \vdots & \vdots & \vdots & \vdots & \ddots & \vdots \\ (t_{a,2})^N & (t_{a,2})^{N-1} & \cdots & t_{a,2} & 0 & 1 & 0 & \cdots & 0 \\ (t_{b,2})^N & (t_{b,2})^{N-1} & \cdots & t_{b,2} & 0 & 1 & 0 & \cdots & 0 \\ (t_{b,2})^N & (t_{c,2})^{N-1} & \cdots & t_{c,2} & 0 & 1 & 0 & \cdots & 0 \\ \vdots & \vdots & \ddots & \vdots & \vdots & \vdots & \vdots & \ddots & \vdots \\ (t_{a,M})^N & (t_{a,M})^{N-1} & \cdots & t_{a,M} & 0 & 0 & 0 & \cdots & 1 \\ (t_{b,M})^N & (t_{b,M})^{N-1} & \cdots & t_{b,M} & 0 & 0 & 0 & \cdots & 1 \\ (t_{c,M})^N & (t_{c,M})^{N-1} & \cdots & t_{c,M} & 0 & 0 & 0 & \cdots & 1 \\ \vdots & \vdots & & \vdots & \vdots & \vdots & \vdots & & \vdots \end{bmatrix} \times \begin{bmatrix} C_N \\ C_{N-1} \\ \vdots \\ C_1 \\ \tau_{01} \\ \tau_{02} \\ \tau_{03} \\ \vdots \\ \tau_{0M} \end{bmatrix} = \begin{bmatrix} \tau_{a,1} \\ \tau_{b,1} \\ \tau_{c,1} \\ \vdots \\ \tau_{a,2} \\ \tau_{b,2} \\ \tau_{c,2} \\ \vdots \\ \tau_{a,M} \\ \tau_{b,M} \\ \tau_{c,M} \\ \vdots \end{bmatrix}. \quad (10)$$

Solving this equation gives the single set of coefficients for the mapping, $C_N, C_{N-1}, ..., C_1$ as well as the $\tau_0$'s. Furthermore, simultaneous fitting to multiple $M$ acquisitions reduces the impact of jitter on the calculated value of the coefficients.

While this transformation could be applied independently for each direction of ECOPS scan, in practice we have found that this is best done using ECOPS measurements which have been "unwrapped", as in Fig. 7 to contain both the forward and backwards sampled signals as they were recorded. That is, the fitting method is applied to both the blue and red points at once, starting with the forward signal and followed by the reverse signal. The small delay between the two directions is the "re-arm" time of the recording instruments, in which no data is collected. This "Full-Cycle" approach more accurately fits the time in between the same landmark features' location in the forward and backwards directions. In other words, the Full-Cycle fitting would correct time-axis locations between 0 and 1 ms in Fig. 7. The true function should then be expected to be periodic with a period equal to that of the frequency modulation. Figure 8 shows the mean-square-error (MSE) of the peak locations in the corrected time-axis results as a function of the fitting polynomial order. These results demonstrate that increasing the order of the fitting polynomial has diminishing returns beyond the 8th order. Therefore, in the subsequent sections, we will use an 8th order polynomial function in Eq. (10).

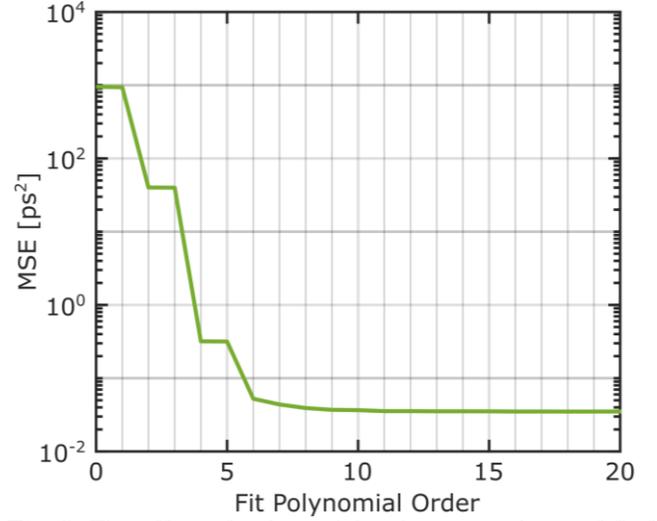

**Fig. 8.** The effect of polynomial order on goodness of fit for time-axis correction as measured by the MSE of ECOPS peak locations compared to ASOPS peak locations. An 8th order polynomial function is selected for subsequent time-axis modeling.

Figure 9 shows the application of the 8th order polynomial time axis correction to Fabry-Perot reflections in Fig. 6. Figure 9(a) shows that this polynomial function, shown by the green dashed line, agrees with the experimental measurements much better in comparison to the theoretical model shown by the black dashed line. In particular, Fig. 9(b) shows that the time-axis error between the theoretical model and the ECOPS measurements can reach several picoseconds in a full-cycle measurement. In Fig. 9(c), however, this error, i.e., $\Delta\tau_{ECOPS}(t)$, the difference between the measured locations and the polynomial model, is approximately uniform and smaller than 1 picosecond. As a result, there is a better match between the ASOPS and ECOPS signals in both directions as shown in Fig. 9(d). Specifically, in Fig. 9(e) the delay between the pulses labeled by $\Delta\tau_{pks}$ in Figs. 6(b) and 9(d) is reduced from 2 and 4 picoseconds in the forward and backwards direction, respectively, to less than 0.3 picosecond in the forward direction (blue) and less than 0.6 picosecond in the backwards direction (red).

When using different ECOPS frequency modulation parameters, i.e., $f_M$ and $\Delta f$ values, following the method described in this section resulted in different 8th order polynomial time-axis corrections, however with similar accuracy in modeling the non-linear time-domain sampling (data not shown). Although results shown in Fig. 9 indicate that the error of time-axis sampling can be markedly reduced, the effect of the residual difference in spectroscopic measurements must be investigated. In the following section, we have used the 8th order polynomial time-axis corrections to model and correct for the non-linear time-axis behavior in evaluating the performance of the new PHASR 2.0 Scanner.

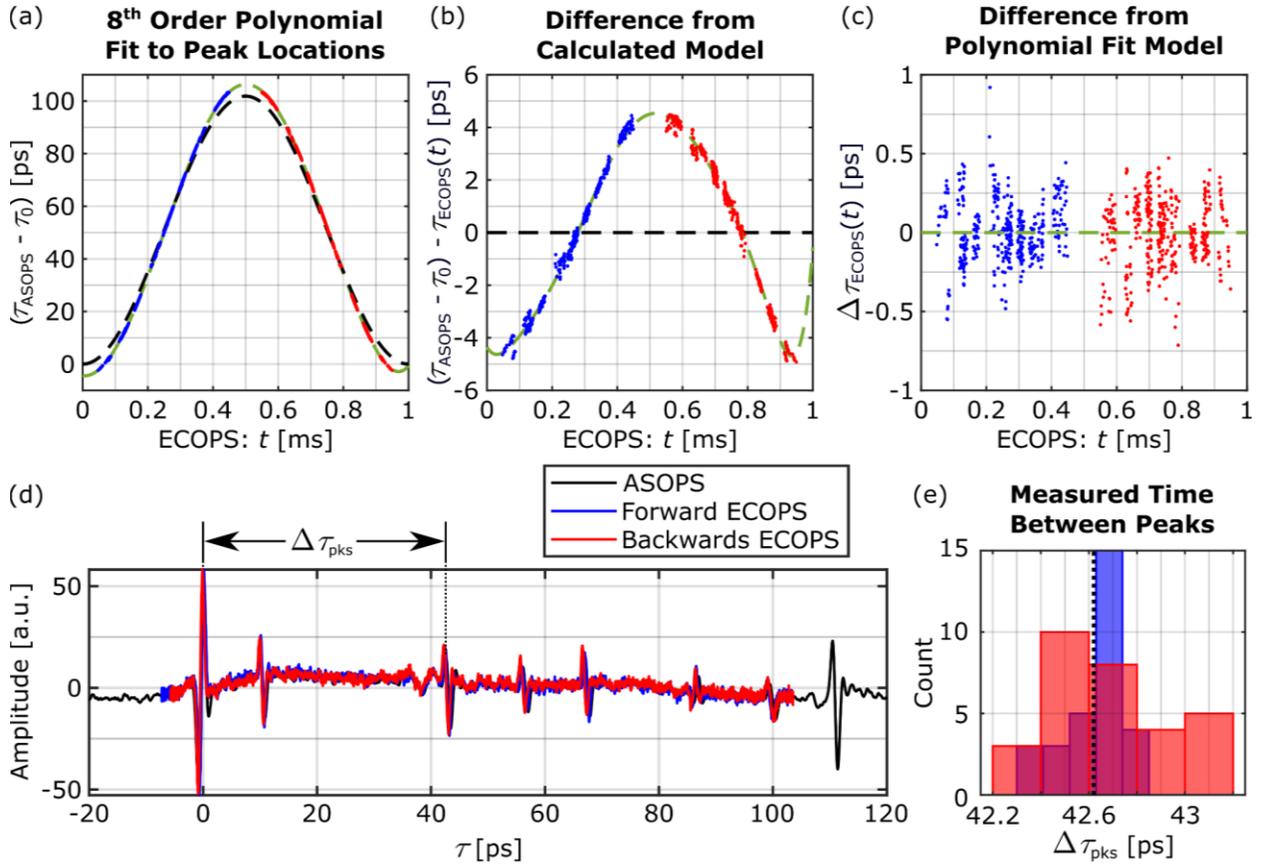

**Fig. 9.** (a) Comparison between the model calculated from Eq. (7) (black dashed line) and polynomial fit (green dashed line) to the actual peak locations (blue and red points). Difference from the calculated model (b) and difference from the polynomial fit (c) highlighting the improved correspondence to ASOPS peak locations. (d) The corrected time domain signal and (e) measurement of $\Delta\tau_{pks}$ demonstrating that after nonlinear time-axis correction the ECOPS traces much more closely match each other and the ASOPS reference. The histogram data are obtained from 30 ECOPS measurements of the sample in Fig. 6.

## IV. INVESTIGATION OF PRACTICAL LIMITS

In order to validate the imaging capabilities of the PHASR Scanner 2.0 and the accuracy of our THz-TDS measurements in the ECOPS mode, we compare measurements against ASOPS data of the same type. In particular, four aspects of the THz-TDS measurements were examined: jitter, dynamic range, usable bandwidth, and spectroscopic accuracy. The first three values were calculated from measurements of a flat mirror while the spectroscopic accuracy was calculated based on the well-studied resonance of lactose at 0.53 THz [23], [48]–[50]. In each case, the performance metric was estimated from 100 independent acquisitions obtained as single point spectroscopy measurements on the sample and without the optional imaging window. These measurements were repeated using different setting values, which affect the acquisition rate: i.e., $f_M$ for ECOPS, as well as the number of time-domain traces averaged per acquisition and $\Delta f$ for both ASOPS and ECOPS. Per the manufacturer's users' manual, the $\Delta f$ values of the ASOPS system can be selected between 1 and 1000 Hz. Since our aim for employing ECOPS measurements is primarily to provide faster acquisition rate than the capabilities of the existing ASOPS system, the examined modulation frequencies were limited to 800 Hz and 1000 Hz. After time-axis correction, a gaussian high-pass filter ($\mu = 0$ THz, $\sigma = 0.05$ THz) was applied to all signals to remove low frequency noise typical of internal reflections within our PHASR Scanner. Table I shows single-shot acquisition time and the maximum THz-TDS sampling range for each of the setting values. Also, Table I includes the dynamic range values, further explored in Sec. IV(B), for 20 and 100 averages of the THz-TDS measurements. For ASOPS measurement, improving acquisition speed by increasing the difference frequency setting results in poorer dynamic range but has no effect on the sampling range. On the other hand, changing the ECOPS modulation settings of $\Delta f$ and $f_M$ have comparatively little effect on the dynamic range of the measurements, but imposes limits on the length of TDS signal, which can be acquired.

### A. Jitter

Consistent timing is vital to time-of-flight and phase-based measurements common to THz-TDS techniques such as material parameter extraction and thickness determination. A

T-TST-REG-04-2022-00056TABLE I
MEASUREMENT CAPABILITIES AT DIFFERENT ASOPS AND ECOPS SETTINGS

| Method | $f_M$ [Hz] | $\Delta f$ [Hz] | Single-Shot Time [ms] | Maximum THz-TDS Sampling Range [ps] | 20-Avg. Dynamic Range [dB] | 100-Avg. Dynamic Range [dB] |
|---|---|---|---|---|---|---|
| ASOPS | 0 | 50 | 20 | 10,000 | 35.4 | 43.2 |
|  | 0 | 100 | 10 | 10,000 | 30.5 | 36.0 |
|  | 0 | 200 | 5 | 10,000 | 21.9 | 28.7 |
|  | 0 | 400 | 2.5 | 10,000 | 14.3 | 21.5 |
|  | 0 | 500 | 2 | 10,000 | 8.7 | 18.7 |
|  | 0 | 1000 | 1 | 10,000 | 4.5 | 9.1 |
| ECOPS | 800 | ±32 | 0.625 | 140 | 34.2 | 42.6 |
|  | 800 | ±42 | 0.625 | 181 | 34.4 | 40.1 |
|  | 1000 | ±15 | 0.5 | 55 | 38.8 | 44.5 |
|  | 1000 | ±25 | 0.5 | 94 | 36.2 | 43.6 |
|  | 1000 | ±32 | 0.5 | 114 | 35.8 | 43.2 |
|  | 1000 | ±35 | 0.5 | 123 | 34.6 | 42.1 |

representative comparison between the ASOPS and ECOPS time-domain measurements obtained at the center of a flat mirror placed at the focal point is shown in Fig. 10(a). We define the jitter by the standard deviation of the time of arrival (ToA) of the reflected pulse as measured by the location of the maximum amplitude in the time-domain. Figure 10(b) shows the performance of both techniques at different settings. Notably, single-shot ASOPS measurements at $\Delta f \geq 400$ Hz suffered from poor SNR such that a typical THz-TDS pulse could not be identified in the recorded trace. For example, single-shot ASOPS acquisitions of $\Delta f = 500$ and 1000 Hz were nearly indistinguishable from noise, requiring special effort to manually locate the correct time window for measurement. Increasing the number of traces averaged per ASOPS acquisition typically decreases the noise, improving precision in calculations involving time of flight or phase measurements. However, this trend reverses for acquisitions which take more than about a second. In contrast, our implementation of ECOPS only improves up to around 20 traces/acquisition. In both cases this indicates that arbitrarily large averaging is inadvisable due to the limits on the stability of the difference frequency, though for ECOPS this is also in part due to the limits of the simple drift compensation model in Sec. III(A). More advanced techniques or implementation of additional hardware such as presented in [43], [46] can improve the large-average performance to better match that of the existing ASOPS system.

## B. Dynamic Range and Usable Bandwidth

In addition to time-resolved measurements, much of the strength of the THz-TDS imaging is due to the ability to measure broadband spectra. Representative frequency domain reference measurements for both ASOPS and ECOPS are shown in Fig. 10(c) along with comparable measurements without the presence of the reference mirror to establish the noise floor. The peak dynamic range, calculated as the maximum ratio of the signal to the noise floor in the frequency domain, is shown in Fig. 10(d). Usable bandwidth is then calculated as the frequency at which the dynamic range first falls to below 3 dB and is plotted in Fig. 10(e). Some ASOPS measurements with high difference frequency and low averaging did not exceed this threshold at any point, resulting in a bandwidth of 0 THz. Since the magnitude of the frequency spectra is not affected by the timing of the pulse, the drift does not affect broadband frequency performance, determined using the magnitude of the Fourier transformation of the TDS pulses, in the same way that it affected the ToA measurements. Thus as expected, increasing the averaging lowers the noise floor, improving both the dynamic range and usable bandwidth of the measurements. The effect of increasing averaging suppresses the noise floor for all settings, resulting in the parallel trends in dynamic range plot. Most notably, ECOPS measurements offer approximately 10-20 dB higher dynamic range as compared to the ASOPS measurements. Furthermore, decreasing the ASOPS difference frequency improves the performance in both measures. ECOPS, which operates with even lower difference frequencies, shows similar capability to that of the best ASOPS setting ($\Delta f = 50$ Hz) when comparing measurements with similar averaging despite the significantly shorter ECOPS measurement times.

The effect of measurement speed on bandwidth is not as easily defined. Two series of water absorption lines beginning at approximately 1.1 and 1.7 THz [51], visible in Fig. 10(c), naturally limit dynamic range in their vicinity and create artificial striation in the measured bandwidth values. We have used a centered moving average filter, with 0.2 THz width applied to the signal spectra, as a simple method to remove the water absorption lines and other spectral fluctuations. Following this step, Figure 10(e) shows that, similar to the dynamic range, the bandwidth of the ASOPS measurements show a marked improvement with decreasing difference frequency. The bandwidth of the ECOPS measurements are higher than all ASOPS settings with similar numbers of averaging. However, the improvement in ECOPS bandwidth with increasing averaging is modest. This behavior is the result of the differing shapes of the spectral density of the noise floor in each technique, as illustrated for instance above 1.7 THz in Fig. 10(c). In general, higher difference frequencies result in a steeper negative slope in the noise floor, while lower difference frequency values produce a flat noise floor.

## C. Spectroscopic Accuracy

To characterize the ability of our modified system to accurately measure frequency spectra, we calculated the measured location of the resonance of lactose, theoretically expected at 0.53 THz. The sample consisted of an approximately 4 mm thick pellet consisting of equal parts by

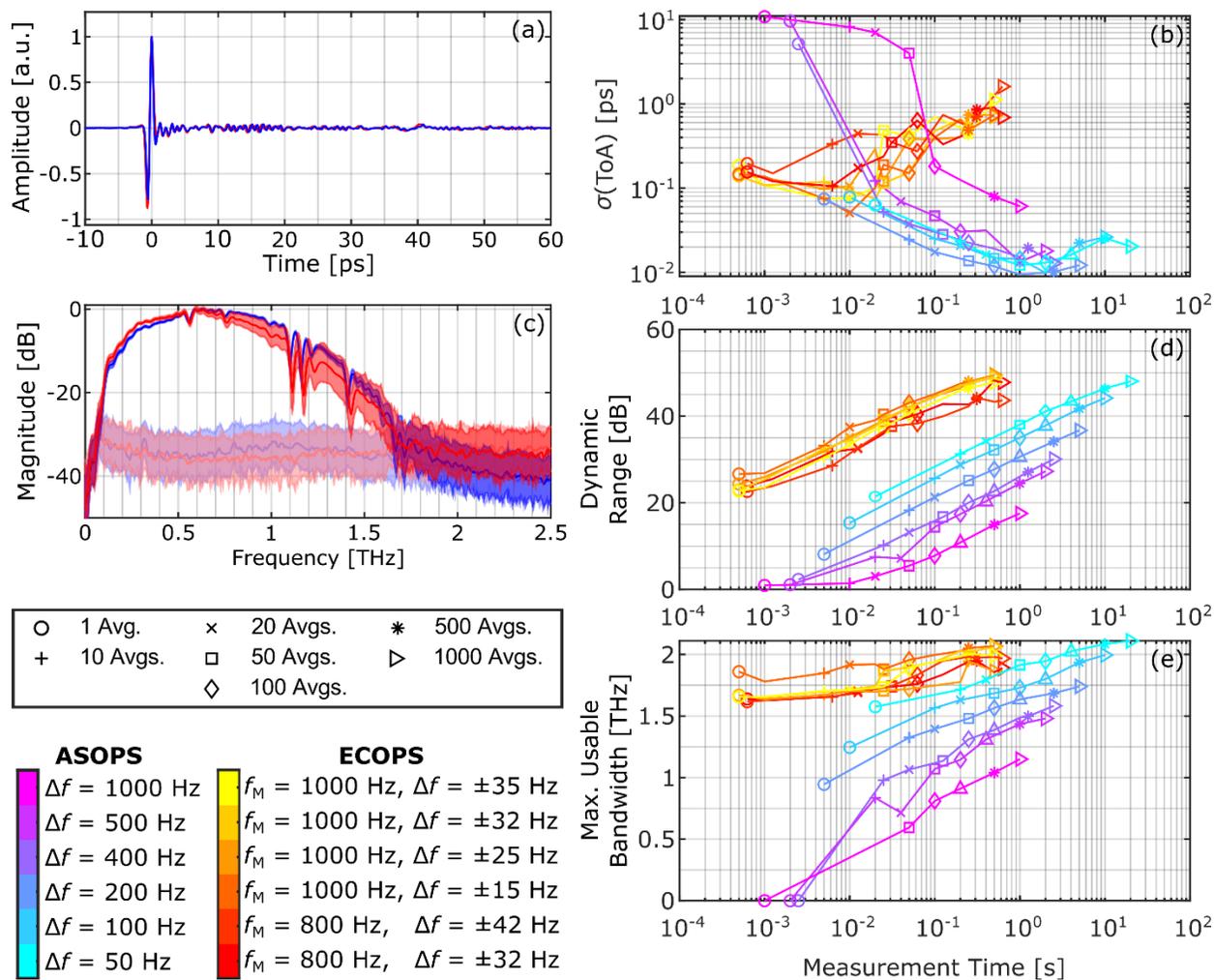

**Fig. 10.** (a) Normalized representative time domain signals reflected from a flat mirror obtained using ASOPS (blue, $\Delta f = 50$ Hz, 20 avg.) and ECOPS (red, $f_M = 1000$ Hz, $\Delta f = \pm 25$ Hz, 20 avg.) methods. (b) Standard deviation of time of arrival (ToA) for all sets of ASOPS and ECOPS acquisitions as a function of measurement time. (c) Fourier transform of the time domain signals in (a) as well as noise measurements in ASOPS (light blue), and ECOPS (light red), acquired with the same parameters. Shaded areas show standard deviation over 100 acquisitions. (d) Maximum dynamic range and (e) maximum usable bandwidth of each set of acquisitions as a function of measurement time.

mass of α-lactose monohydrate and high-density polyethylene (for binding). The two components were mixed as powders with mortar and pestle and then compressed for 3 hours. The sample was placed on a mirror and the reflection from the back surface (that is, the signal which has passed through twice the sample thickness) was captured. The location of the resonance was then determined by finding the location of the minimum spectral amplitude in the area between 0.45 and 0.65 THz. This test provides additional insight into how well the nonlinear time-axis sampling correction performs over large sections of the signal, as incorrect scaling will lead to frequency shifts. Figures 11(a)-(b) show the distribution of the resonance locations using a selection of ASOPS and ECOPS settings after averaging 20 independent traces. For ECOPS measurements, the distribution of the resonances calculated for each direction using Eq. (7) (i.e., without time-axis correction) are also shown in red and blue. It can be seen that the precision of ASOPS measurements improves as the difference frequency is lowered, however the accuracy of ECOPS measurements remains higher than ASOPS and independent of ECOPS settings after our proposed time-axis correction method. Figure 11(c), for example, compares the spectral location of lactose's resonance for ASOPS measurements marked with the light blue box in Fig. 11(a) ($\Delta f = 100$ Hz) with ECOPS measurements selected by the orange box in Fig. 11(b) ($f_M = 1000$ Hz, $\Delta f = \pm 32$ Hz) before and after the nonlinear time-axis correction. Overall, the time-axis corrected ECOPS results perform better than even the $\Delta f = 50$ Hz ASOPS measurement, consistent with the trend according to difference frequency described previously. As shown in Fig. 11(d), the precision of this measurement improves as increasing averaging reduces the noise. This precision, as measured by standard deviation of the absorption peak location, is plotted for all measurement settings. While the variation decreases with increasing averaging as expected for both techniques, the standard deviation of the ASOPS results is nearly an order of magnitude higher than the corresponding

T-TST-REG-04-2022-00056                                                                                                                                                                                                                                                  12

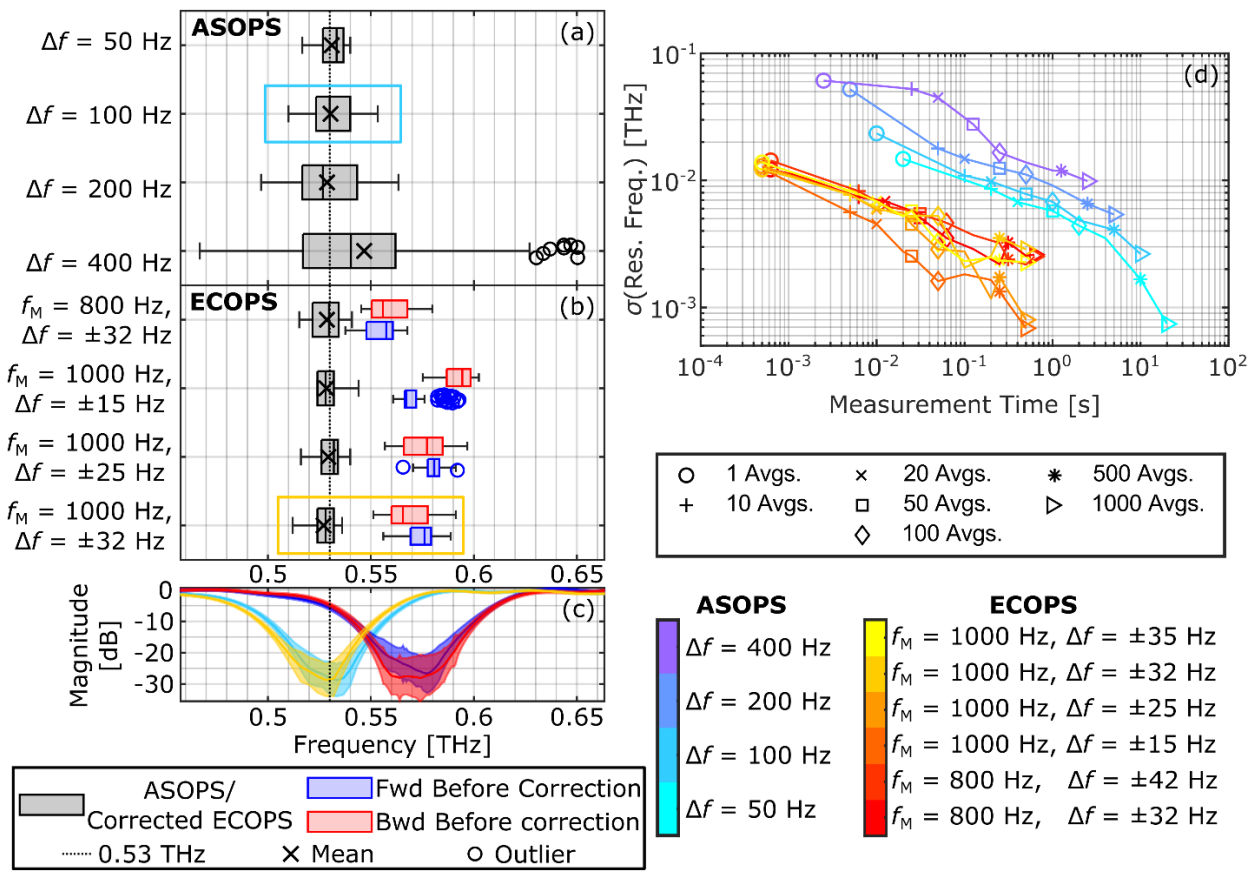

**Fig. 11.** Distribution of measured lactose resonance. (a) ASOPS measurements and (b) ECOPS measurements including values calculated from forward (blue) and backwards (red) directions using Eq. (7) as well as time-axis corrected (black). (c) Frequency domain plot of select ASOPS ($\Delta f = 100$ Hz, 20 avg., light blue box in (a)) and ECOPS ($f_M = 1000$ Hz, $\Delta f = \pm 32$ Hz, 20 avg, blue and red for ECOPS forward and backwards directions before time-axis correction, respectively, and orange for after correction). Area demonstrates mean ± standard deviation among 100 acquisitions (50 each of forward and backwards for ECOPS signals). Light blue- and orange-colored boxes in (a) and (b) indicate the corresponding boxplots. The resonance of lactose at 0.53 THz in time-axis corrected ECOPS measurements (orange) overlap closely with ASOPS results (light blue). (d) Standard deviation of measured resonance location for each set of acquisitions. ECOPS measurements shown only after time-axis correction.

ECOPS measurements which have the same acquisition time. Caution should be taken not to interpret these values as the frequency resolution of the signals—which is determined by the signal length upon which the Fourier transform is applied—but rather as a measure of the repeatability of measurements. For example, if the extracted values of the resonance location are split between a relatively few number of close frequency bins, the calculated standard deviation can be smaller than the frequency resolution for that measurement setting.

## V. FAST ACQUISITION DEMONSTRATION

Finally, we show the overall improvement of our PHASR Scanner 2.0 by demonstrating its scanning capabilities in situ. A video demonstrating a scanning time of approximately 8 seconds over a 27×27 mm² FOV with 1 mm pixel sizes, (i.e., a 729-pixel image) is presented in the supplemental materials. The scan was acquired at $f_M = 1000$ Hz, $\Delta f = \pm 32$ Hz, for a 2000 THz-TDs trace/s acquisition rate. The acrylic target and peak-to-peak amplitude image of the scan is shown in Fig. 12(a) and (b), respectively. Each pixel represented the average of 10 time-domain traces. To ensure that the pixels conform to a grid, each line of the scan consisted of an acceleration period, a constant speed section covering the FOV, and then a deceleration period and movement to the next line. Data was acquired during the constant speed section without pausing the beam-steering for each pixel. The acceleration, deceleration, and line step periods added an additional overhead time of 154 ms/line or 4.0 seconds for an entire image, during which THz traces were not used.

A 1951 USAF Resolution Test Target provides a demonstration case for the full field of view of the scanner. The area containing elements 4-6 of group -2 (line widths ranging from 1.41 to 1.12 mm) and the resulting THz peak-to-peak image are shown in Fig. 12(c) and (d), respectively. The circular area of the ECOPS image clearly shows the boundaries of the lens area.

## V. CONCLUSION

Our first effort at a portable handheld scanner, the PHASR Scanner 1.0, addressed many of the problems present in current applications of THz-TDS imaging but it also had clear limitations in field of view and scanning speed. Implementing a heliostat gimbal geometry drastically reduced the inherent distortion from the scanning system and improved the scanning range from 12×19 mm² to 40×27 mm². This is combined with small modifications to the commercial ASOPS system, which allowed ECOPS operation of up to 2000 trace/s measurement rate. To implement this change, we used the existing ASOPS hardware, though specific attention is required to reduce signal drift and non-linear time-axis sampling inherent to this upgrade. In particular, we used a state model to make real-time corrections to the time-window and a polynomial time-axis calibration to an ASOPS measurement based on Fabry-Perot reflections for accurate time-axis scaling. The resulting polynomial fit can then be used for further measurements within that session or until the ECOPS modulation parameters are changed. We demonstrated the performance metric of the new ECOPS-based PHASR 2.0 Scanner. We show that we can use the ECOPS mode to take measurements with similar or better frequency-domain performance in significantly less time. These improvements make the PHASR Scanner 2.0 much more practical to use in scenarios such as biomedical imaging where scanning field of view and scanning speed significantly affect the patient experience. Our future work with the PHASR 2.0 scanner is intended to demonstrate its ability "in the field" for clinical and industrial applications. Results of these studies have already been implemented to extend the FOV and speed of our PHASR 1.0 Scanner for imaging large burns with 1" diameter in several preclinical in vivo studies [52]–[54]. Furthermore, recent results have demonstrated the value of polarization-sensitive THz measurement of biological samples [55], including skin [56]. This motivates successive designs which will develop our work on THz polarimetry [57] with the goal of providing fast, portable terahertz ellipsometry.

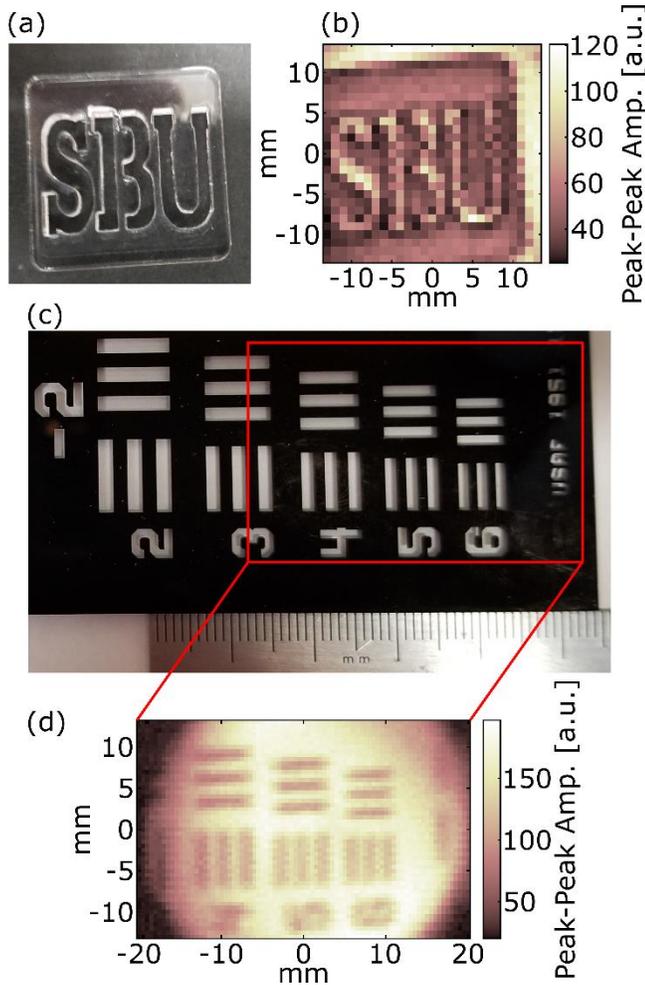

**Fig. 12.** Demonstration images. (a) Photograph and (b) Peak-to-peak image of the acrylic SBU target acquired during the 8-second ECOPS THz-TDS scan presented in the supplemental materials. (c) Visual image of group -2 of a 1951 USAF Resolution Test Target and (d) the corresponding THz peak-to-peak image from an ECOPS scan. The full FOV (corresponding to the red box in (c)) of the scanner is shown. The vertical direction is limited by the range of the goniometer while the horizontal range is limited by the aperture of the lens (circular profile).

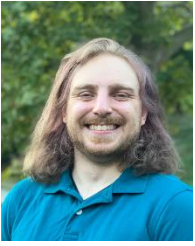 **Zachery B. Harris** received the B.S. degree in physics from the University of Washington, Seattle, WA in 2017.
He is currently working as a Research Technician at Stony Brook University. His research interests include terahertz imaging and spectroscopy, terahertz signal processing, nondestructive evaluation, and medical imaging.

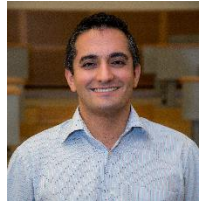 **M. Hassan Arbab** received the B.S. degree in electrical engineering from Shahid Beheshti University, Tehran, Iran, in 2004 and the M.S. and Dual Ph.D. degrees in electrical engineering and nanotechnology from the University of Washington, Seattle, WA in 2008 and 2012, respectively.

From 2012 to 2016 he was a Postdoctoral Research Associate and a Senior Research Scientist with the Applied Physics Laboratory at the University of Washington. Since 2016, he has been an Assistant Professor with the Biomedical Engineering Department, Stony Brook University, Stony Brook, NY. His research interests include terahertz science and technology, ultrafast and nonlinear optics, signal and image processing, machine learning and biomedical applications of terahertz spectroscopy.

Dr. Arbab is a member of the American Physical Society, the Optical Society of America, The International Society for Optical Engineering and the Biomedical Engineering Society.